\documentclass[conference, letterpaper]{IEEEtran}
\usepackage[pdftex]{graphicx}
\usepackage[caption=false]{subfig}
\usepackage{booktabs}
\usepackage{array}
\usepackage{colortbl}
\usepackage{color}
\usepackage{float}
\usepackage[hyphens]{url}
\usepackage{threeparttable}
\begin{document}
\title{The use of Charts, Pivot Tables, and Array Formulas in two Popular Spreadsheet Corpora}
\author{\IEEEauthorblockN{Bas Jansen}
\IEEEauthorblockA{Delft University of Technology\\
Email: b.jansen@tudelft.nl}
\and
\IEEEauthorblockN{Felienne Hermans}
\IEEEauthorblockA{Delft University of Technology\\
Email: f.f.j.hermans@tudelft.nl}
}
\maketitle
\begin{abstract}
The use of spreadsheets in industry is widespread. Companies base decisions on information coming from spreadsheets. Unfortunately, spreadsheets are error-prone and this increases the risk that companies base their decisions on inaccurate information, which can lead to incorrect decisions and loss of money. In general, spreadsheet research is aimed to reduce the error-proneness of spreadsheets. Most research is concentrated on the use of formulas. However, there are other constructions in spreadsheets, like charts, pivot tables, and array formulas, that are also used to present decision support information to the user. There is almost no research about how these constructions are used. To improve spreadsheet quality it is important to understand how spreadsheets are used and to obtain a complete understanding, the use of charts, pivot tables, and array formulas should be included in research. In this paper, we analyze two popular spreadsheet corpora: Enron and EUSES on the use of the aforementioned constructions.
\end{abstract}
\section{Introduction} 
\label{sec:introduction}
The use of spreadsheets in industry is widespread. It is estimated that 90\% of all analysts use spreadsheets for their calculations \cite{winston2001executive} and 95\% of US firms use spreadsheets in some form of financial reporting \cite{panko2008sarbanes}. This reporting and the calculations of annalists in spreadsheets are used by companies to make decisions. Unfortunately, spreadsheets are known to be error-prone \cite{panko1998we}. This increases the risk that companies make decisions that are based on inaccurate information which can lead to incorrect decisions and loss of money\footnote{http://www.eusprig.org/horror-stories.htm}.

A substantial part of current spreadsheet research is focused on improving the quality of spreadsheets by applying software engineering methods to them like testing \cite{rothermel2000wysiwyt} \cite{roy2017spreadsheet}, reverse engineering \cite{Hermans2010} \cite{cunha2010automatically}, code smells \cite{hermans2014detecting} \cite{cunha2012towards}, and refactoring \cite{hermans2014bumblebee} \cite{badame2012refactoring}. This research has in common that it concentrates on spreadsheet formulas. However, there are other constructions in spreadsheets like charts, pivot tables, and array formulas, that are also used to present decision support information to the user of the spreadsheet. We do not know how widespread the use of these constructions are and if they are somehow related to the error-proneness of spreadsheets. There is almost no research on this topic. The only study we found is the paper of Fisher and Rothermel in which they introduce the EUSES corpus. They counted the number of spreadsheets that contained charts.

To improve the overall quality of spreadsheets, we need to understand how spreadsheets are used in industry, and to obtain a complete understanding we should include constructions like charts, pivot table and array formulas in our research. Therefore, in this short paper, we analyze two popular spreadsheet corpora: Enron \cite{hermans2015enron} and EUSES \cite{fisher2005euses} on the use of charts, pivot tables, and array formulas.

The contributions of this paper are: 1) a tool that is capable of analyzing properties of charts, pivot tables, and array formulas in Excel files and 2) an analysis of the use of charts, pivot tables, and array formulas in the EUSES and Enron corpora.

We organize the remainder of this paper in the following way. In the next section, we describe the approach that was used to analyze the spreadsheets. In Section \ref{sec:results} we present the results of the analysis. Related work is discussed in Section \ref{sec:related_work} and we end the paper with the concluding remarks in Section \ref{sec:concluding_remarks}.

\section{Approach and Tool} 
\label{sec:approach}
In this study we use two popular spreadsheet corpora: EUSES \cite{fisher2005euses} and Enron \cite{hermans2015enron}. In previous work \cite{jansen2015corpora}, \cite{Hermans:2012aa}, \cite{Jansen:2015aa} we used the Spreadsheet Scantool, developed at Delft University of Technology. It utilizes the Gembox library to read spreadsheets. Unfortunately, Gembox is not able to analyze charts, pivot tables or array formulas. As a result, we developed a new Analyzer in Visual Basic for Applications (VBA) that is able to access the full Excel object model. We made the VBA code together with instructions about how to use it, available on Github\footnote{\url{https://github.com/HeerBommel/SpreadsheetExplorer.git}}. With the Analyzer, we collect several metrics about the use of charts, pivot tables, and array formulas in spreadsheets in the aforementioned corpora.

For charts, we counted the number of spreadsheets that contained at least one chart and the total number of charts in the corpus. For each chart, we determined the chart type. In Excel there are 73 different chart types\footnote{https://msdn.microsoft.com/en-us/vba/excel-vba/articles/xlcharttype-enumeration-excel}. However, most chart types are a variation of their base type. For example, the `radar', `filled radar' and `radar with data markers' all belong to the category of radar charts. In our analysis, we mapped the actual chart type to its category.

Also for pivot tables, we counted the number of spreadsheets that contained at least one pivot table. Next, for each pivot table we analyzed:
\begin{itemize}
    \item the size: by counting the number of rows and columns of the underlying dataset
    \item the number of calculated fields: calculated fields are formulas that can refer to other fields in the pivot table
    \item the number of calculated items: calculated items are formulas that can refer to other items within a specific pivot field
    \item the aggregation functions used
    \item the number of pivot tables per worksheet
\end{itemize}

With respect to array formulas, we counted the number of spreadsheets that contained at least one array formula and we analyzed all array formulas to obtain a better understanding of why they are used.

\section{Results} 
\label{sec:results}
In this section, we will present the results of our analysis. The spreadsheets from both the Enron and the EUSES corpus were scanned for the usage of charts, pivot tables, and array formulas. For each construct, a more detailed analysis was executed to gain more insight into how they were used.

\subsection{Charts} 
\label{sub:charts}

Table \ref{tab:NoSpreadsheetsAnalyzed} shows the total number of spreadsheets we have analyzed. Some of the files were password protected, corrupted or otherwise unreadable and we have excluded them from the study (25 files in EUSES and 130 in the Enron dataset). Charts are not rare but the majority of spreadsheets do not contain any charts. They occur more in the Enron spreadsheets. This could be an indication that charts are used more frequently in an industrial setting.

\begin{table}[ht]
\caption{Number of spreadsheets with and without charts}\label{tab:NoSpreadsheetsAnalyzed}
\centering
\begin{tabular}{lrrrr}
\hline
& EUSES & \% & Enron & \% \\
\hline
Charts & 340 & 8\% & 1,721 & 11\% \\
No Charts & 4,133 & 92\% & 14,078 & 89\% \\
\hline
Total & 4,473 & 100\% & 15,799 & 100\% \\
\hline
\end{tabular}
\end{table}

Within Excel there are two ways to create a chart: 1) the chart is created as a special type of worksheet, this is called a Chart Sheet or 2) the chart is embedded on an existing worksheet. Table \ref{tab:ChartSheetsVsEmbedded} presents the occurrence of the two types in both corpora. When Fisher and Rothermel introduced the EUSES corpus, they stated that 105 of the 4,498 spreadsheets contained charts \cite{fisher2005euses}. We found a different number. According to our analysis, there are 340 spreadsheets that contain charts. The difference is caused by the two different ways a chart can be created. Fisher and Rothermel only counted the chart sheets, while our analysis also included the embedded charts.

\begin{table}[ht]
\caption{The use of Chart Sheets vs Embedded Charts}\label{tab:ChartSheetsVsEmbedded}
\centering
\begin{tabular}{l|rr|rr}
\hline
& \multicolumn{2}{c|}{EUSES} & \multicolumn{2}{c}{Enron} \\
Type & \# Charts & \% & \# Charts & \% \\
\hline
Sheet & 355 & 25\% & 1,149 & 13\% \\
Embedded & 1,090 & 75\% & 7,686 & 87\% \\
Total & 1,445 & 100\% & 8,835 & 100\% \\
\hline
\end{tabular}
\end{table}

\begin{figure}[ht]
    \centering
    \includegraphics[clip,width=\columnwidth]{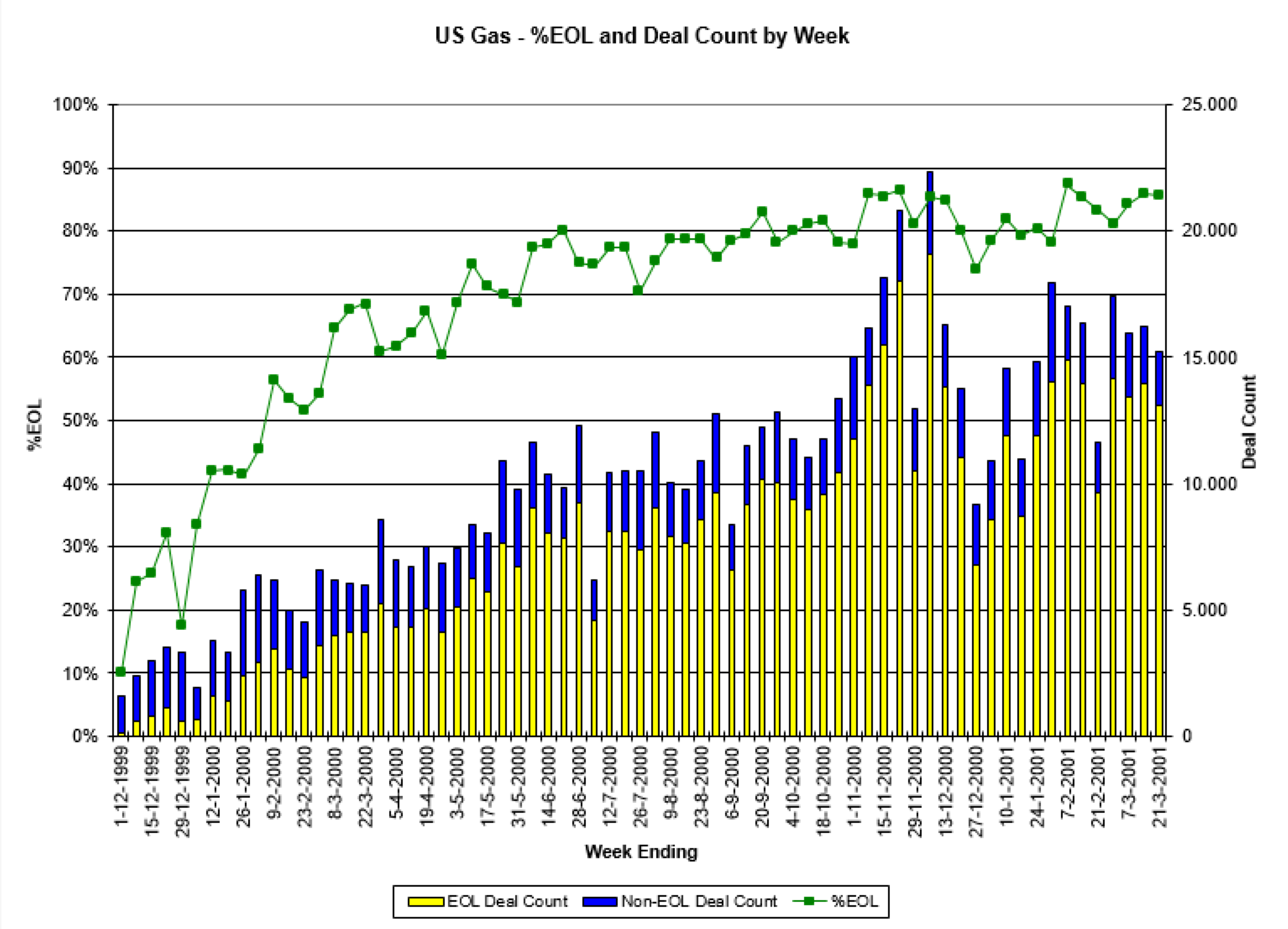}
    \caption{Mixed chart type (Source: Enron Corpus scott\_neal\_\_38010\_\_Helplist.xls, sheet US Gas)}
    \label{fig:ExampleMixedTypeChart}
\end{figure}

Tables \ref{tab:ChartTypesEUSES} and \ref{tab:ChartTypesEnron} show the different chart types that were used in both corpora. The most frequently used type is the column chart. Other popular types in both corpora are the line chart and the pie chart. Notable differences between the two corpora are the occurrence of the mixed and scatter chart types. The scatter chart is used frequently in the EUSES corpus (22\%), but less frequently in the Enron set (2\%). The opposite is true for the mixed chart type: it is frequently used within Enron but hardly in the EUSES set. A mixed chart type means that either two y-axes are used or there are two or more data series that use a different chart type (for example the combination of a line and a column chart), see Figure \ref{fig:ExampleMixedTypeChart} for an example.      

\begin{table}[ht]
\caption{Overview of used chart types in EUSES}\label{tab:ChartTypesEUSES}
\centering
\begin{tabular}{rlrrrlrr}
\hline
Rank & Chart Type & \# Charts & \% \\
\hline
1 & Column & 515 & 35.6\% \\
2 & Scatter & 322 & 22.3\% \\
3 & Line & 258 & 17.9\% \\
4 & Pie & 139 & 9.6\% \\
5 & Bar & 125 & 8.7\% \\
6 & Mixed & 55 & 3.8\% \\
7 & Surface & 15 & 1.0\% \\
8 & Area & 13 & 0.9\% \\
9 & Stock & 2 & 0.1\% \\
10 & Radar & 1 & 0.1\% \\
\hline
\end{tabular}
\end{table}

\begin{table}[ht]
\caption{Overview of used chart types in Enron}\label{tab:ChartTypesEnron}
\centering
\begin{tabular}{rlrr}
\hline
Rank & Chart Type & \# Charts & \% \\
\hline
1 & Column & 4,253 & 48.1\% \\
2 & Line & 2,815 & 31.9\% \\
3 & Mixed & 866 & 9.8\% \\
4 & Pie & 649 & 7.3\% \\
5 & Scatter & 168 & 1.9\% \\
6 & Area & 67 & 0.8\% \\
7 & Bar & 11 & 0.1\% \\
8 & Surface & 6 & 0.1\% \\
\hline
\end{tabular}
\end{table}

\subsection{Pivot Tables} 
\label{sub:pivot_tables}

Table \ref{tab:NoSpreadsheetsWithPivot} shows the number of spreadsheets in both corpora that contain pivot tables. It shows that pivot tables are used very rarely but that they are used more frequently in the industrial Enron set than in the EUSES spreadsheets. 

\begin{table}
\caption{Number of Spreadsheets with Pivot Tables}\label{tab:NoSpreadsheetsWithPivot}
\centering
\begin{tabular}{lrrrr}
\hline
& EUSES & \% & Enron & \% \\
\hline
Pivot & 19 & 0.4\% & 244 & 1.5\% \\
No Pivot & 4,454 & 99.6\% & 15,555 & 98.5\% \\
\hline
Total & 4,473 & 100.0\% & 15,799 & 100.0\% \\
\hline
\end{tabular}
\end{table}
\begin{table}
\hfill
\caption{Overview of Pivot Table Metrics}\label{tab:PivotMetrics}
\centering
\begin{tabular}{lrrrr}
\hline
& avg. \# & avg. \# & \# calc. & \# calc. \\
Corpus & records & fields & items & fields \\
\hline
EUSES & 3,738 & 13 & 3 & 0 \\
Enron & 4,073 & 28 & 0 & 0 \\
\hline
\end{tabular}
\end{table}

Pivot tables can be used to analyze large datasets. Therefore, we analyzed the average size of the data tables behind the pivot tables (Table \ref{tab:PivotMetrics}). The maximum number of records we found is equal to the maximum number of rows up to Excel 2003 (65,535 rows). The average number of records in both datasets are comparable, but in the Enron sheets, the average number of columns (fields) in these data sets is larger. Within a pivot table, it is possible to create calculated fields or items based on other data in the table. From Table \ref{tab:PivotMetrics} we can observe that this functionality is hardly used. 

The main functionality that a pivot table provides is the possibility to summarize large data sets with aggregate functions. We analyzed the use of these functions and found that, in both corpora, in more than 85\% of the cases, either the function Sum or Count was used to summarize the data.

When a pivot table is defined in Excel, by default it is created on a new worksheet. However, it is possible to create more than one pivot table on a worksheet and based on our analysis we can conclude that users tend to do so. The average number of pivot tables per worksheet ranges from 1.3 (EUSES) to 1.6 (Enron) and we found a maximum of nine pivot tables on one worksheet.

\subsection{Array Formulas} 
\label{sub:array_formulas}
Formulas in Excel always return a single value. However, with an array formula\footnote{\url{https://support.office.com/en-us/article/Guidelines-and-examples-of-array-formulas-7d94a64e-3ff3-4686-9372-ecfd5caa57c7?ui=en-US&rs=en-US&ad=US}}, it is possible to perform multiple calculations on one or more items of an array and return either a single value or multiple values. The formulas have to be entered with a special key combination (ctrl + shift + enter) and most users are unaware of their existence. Nevertheless, we checked in both corpora for array formulas and to our surprise, we found that their occurrence is similar to that of pivot tables. About 1.2\% of the spreadsheets (in both corpora) contains array formulas.

In the 250 spreadsheets that contained one or more array formulas, we found a total of 3,965 unique array formulas. We have analyzed these formulas to obtain a better understanding of why array formulas are used. Table \ref{tab:ArrayUseCases} summarizes the results.

\begin{table}[ht]
\hfill
\caption{Use cases for Array Formulas}\label{tab:ArrayUseCases}
\centering
\begin{tabular}{lrrrr}
\hline
Use case & \# Formulas & \% of Formulas \\
\hline
Multiple criteria aggregation & 2,398 & 60.5\% \\
Array functions & 690 & 17.4\% \\
Sumproduct & 455 & 11.5\% \\
User defined functions & 264 & 6.7\% \\
Unnecessary or erroneous & 66 & 1.7\% \\
Repeat block & 47 & 1.2\% \\
TABLE function & 45 & 1.1\% \\
\hline
Total & 3,965 & 100.0\% \\
\hline
\end{tabular}
\end{table}

\begin{figure}[ht]
    \centering
    \includegraphics[scale=0.30]{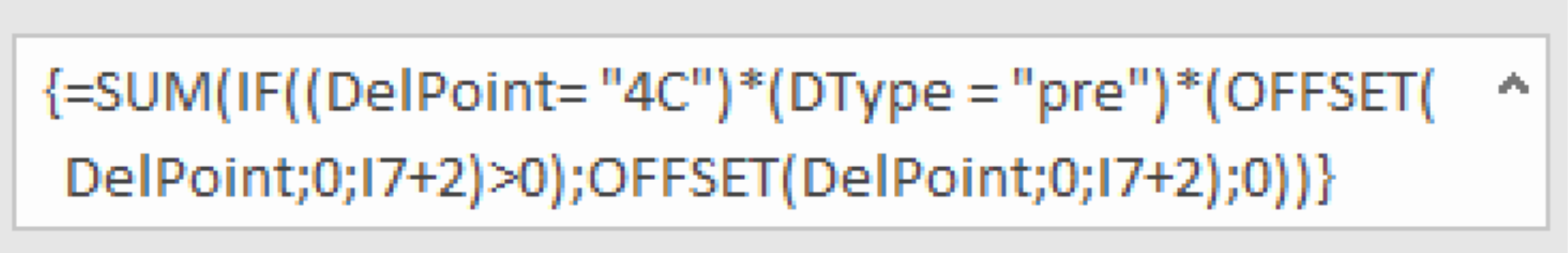}
    \caption{Array formula for a multiple criteria SUM (Source: Enron Corpus eric\_linder\_\_9655\_\_4\_02act.xlsx, sheet Preschedule, cell I30)}
    \label{fig:ExampleArray}
\end{figure}

In over 60\% of the cases, array formulas are used to calculate aggregated values with functions like SUM, MIN, and AVERAGE, based on multiple criteria. An example of such a formula is shown in Figure \ref{fig:ExampleArray}.

The curly brackets indicate that the formula is an array formula. This specific use case is related to the age of both the Enron and the EUSES corpus. At the time the spreadsheets in the Enron and EUSES corpus were created, there were no functions available for conditional aggregation of values and it could only be accomplished by using an array formula or the SUMPRODUCT function. However, as from Excel 2010, Microsoft has added dedicated functions like SUMIFS, AVERAGEIFS, MINIFS, etc. for these type of calculations.

Another important use case for array formulas are the special array functions in Excel, like TRANSPOSE, MMULT, LINEST, FREQUENCY, and MINVERSE. Instead of returning a single value, these functions return an array as result and can only be used in an array formula.

Sometimes an array formula is entered as a multi-cell array formula. In such a case a group of cells gets exactly the same array formula and this formula can only be edited when the whole group is selected. The formula will calculate the result for each item in the array of cells. We grouped this use case under the category repeat block.

Other use cases we found were user-defined functions that are used as array formulas and what-if analyses with a data table (TABLE function \footnote{https://support.office.com/en-us/article/calculate-multiple-results-by-using-a-data-table-e95e2487-6ca6-4413-ad12-77542a5ea50b}). Finally, we encountered a set of array formulas that were either incorrect or unnecessary. The latter meaning that the same formula would also have worked without the special array syntax.

\section{Related Work} 
\label{sec:related_work}

Most related to our work are the papers introducing the corpora that we have analyzed, Euses~\cite{fisher2005euses} and Enron~\cite{hermans2015enron}. While the paper introducing EUSES describes statistics on charts too the paper on Enron does not. Another spreadsheet corpus is FUSE \cite{barik2015f}. This corpus consists of almost 250,000 spreadsheets that were extracted from a public web archive with over 26 billion pages. We preferred the Enron corpus over the FUSE corpus because the Enron spreadsheets were used in an industrial setting.

Besides those two corpora, there are other smaller corpora: Two prominent corpora are the Galumpke and Wall corpora, containing 82 and 150 spreadsheets respectively, both derived from classroom experiments~\cite{panko2000two}. Powell and colleagues survey other corpora used in field audits, each audit examining between 1 and 30 spreadsheets~\cite{powell2008critical}. To our knowledge, none of these corpora are publicly available. 

Furthermore, there are papers on spreadsheet metrics, which also measure properties of spreadsheets. In 2004, Bregar published a paper presenting a list of spreadsheet metrics based on software metrics~\cite{breg2004}. He, however, does not provide any justification of the metrics, nor did he present an evaluation.


\section{Concluding Remarks} 
\label{sec:concluding_remarks}
In this paper, we analyzed two popular spreadsheet corpora and focussed on the use of charts, pivot tables, and array formulas. Although the spreadsheets in both corpora are more than ten years old, we believe they still offer a valuable insight into how they are used. Especially because Microsoft did not make a lot of changes with respect to charts, pivot tables, and array formulas.

We found that charts are used in about 10\% of the spreadsheets. The most commonly used chart types are Column, Line, and Pie. Pivot tables are much rarer and only found in about 1\% of the spreadsheets. In more than 85\% of the cases, the data in the pivot table is summarized with the aggregate functions \emph{Sum} and \emph{Count}. Finally, to our surprise, the complex array formulas are used as frequently as pivot tables.

Overall the use of charts, pivot tables, and array formulas in spreadsheets is limited. Still, they can have an impact on the error-proneness of spreadsheets. Especially pivot tables and array formulas can introduce new types of errors. A well-known problem with pivot tables is that they are not automatically refreshed when the underlying data changes, increasing the risk that the user is analyzing outdated data. Array formulas are difficult to understand and not very well known by the majority of the spreadsheet users. Because of their complexity, it is easy to make errors in these formulas. The information presented in the charts is coming from the underlying data. Errors in this data will lead to errors in charts. However, charts also can introduce their own errors. For example, Excel will in some cases automatically let the Y axis not start at zero, which could misrepresent the underlying data.


\end{document}